\documentclass{PoS}
\usepackage{axodraw}
\def\be{\begin{equation}}
\def\ee{\end{equation}}

\title{Universality and massive excitations in 3$d$ 3-state Potts model}

\ShortTitle{Universality and massive excitations in 3$d$ 3-state Potts model}

\author{R. Falcone$^a$, R. Fiore$^a$, \speaker{M. Gravina}$^a$ and A. Papa$^a$\\ 
        \llap{$^a$}Dipartimento di Fisica, Universit\`a della Calabria,\\
        and Istituto Nazionale di Fisica Nucleare, Gruppo collegato di Cosenza\\
        I--87036 Arcavacata di Rende, Cosenza, Italy\\
        E-mail: \email{rfalcone,fiore,gravina,papa@cs.infn.it} }

\abstract{The mass spectrum of the 3$d$ 3-state Potts model is considered in the broken phase 
(a) near the second order Ising critical point in the temperature-magnetic field plane
and (b) near the weakly first order transition point at zero magnetic field.
In the case (a), the mass spectrum is compared with the prediction from universality of mass ratios in 
the 3$d$ Ising class; in the case (b) a mass ratio is determined to be compared with 
the corresponding one in the spectrum of screening masses of the (3+1)$d$ SU(3) pure gauge
theory at finite temperature in the deconfined phase near the transition.}

\FullConference{The XXV International Symposium on Lattice Field Theory\\
		 July 30-4 August 2007\\
		 Regensburg, Germany}

\begin{document}

\section{Introduction}

The Svetitsky-Yaffe conjecture~\cite{Svetitsky:1982gs} establishes that $(d+1)$-dimensional 
SU(N) pure gauge theories at finite temperature, which undergo a confinement/deconfinement phase 
transition associated with the breaking of the center of the gauge group Z(N)~\cite{Polyakov:vu,Susskind:up}, 
are in the same universality class of $d$-dimensional Z(N) spin models, which undergo an 
order/disorder transition. This implies that the gauge theory and the spin model present the 
same universal quantities (critical indices, amplitude ratios, ...) at criticality. This 
has been verified in the case of a second order transition by numerical investigation in several
papers (see, for instance, Refs.~\cite{Gliozzi:1997yc} and, for a review, Ref.~\cite{Pelissetto-Vicari:2002}).
Moreover recently it has been conjectured that also the ratios of massive excitations 
in the broken phase are universal~\cite{Caselle:1999tm} and a numerical evidence of this 
conjecture has been given in the case of 3$d$ Ising universality 
class~\cite{Caselle:2001im,Fiore:2002fj,Fiore-Papa-Provero-2003}.

\begin{figure}[b]
\centering
\label{phase_diag_xitau}
\begin{picture}(250,160)(0,0)

\LongArrow(0,10)(250,10)
\LongArrow(0,10)(0,160)
\SetWidth{2.5}
\Line(100,10)(250,10)
\CArc(300,100)(219.3,190,204)
\CCirc(84,62){2}{1}{1}
\SetWidth{0.5}
\LongArrow(70,140)(95,0)
\LongArrow(150,75)(10,49)

\Text(250,0)[r]{$\beta$}
\Text(-10,160)[r]{$h$}
\Text(110,20)[c]{$\beta_t$}
\Text(60,75)[c]{$(\tau_c,\xi_c)$}
\Text(10,60)[c]{$\xi$}
\Text(85,0)[c]{$\tau$}

\end{picture}
\caption[]{Qualitative phase diagram of the 3$d$ 3-states Potts model
in the $(\beta,h)$-plane: the solid line in bold is the line of first order phase 
transitions. $\beta_t$ is the order/disorder inverse transition temperature at zero 
magnetic field;  $\xi$ and $\tau$ are the symmetry-breaking and the temperature parameters 
of the Ising theory~(\ref{ising}) and $(\tau_c,\xi_c)$ is the endpoint in the 3$d$ Ising class.}
\end{figure}
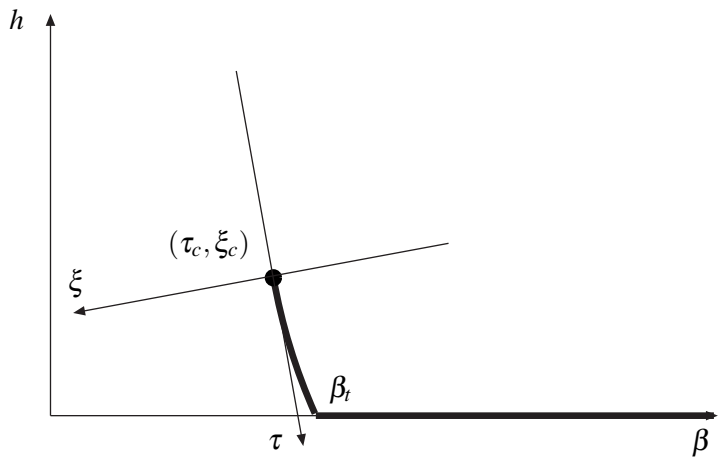

In this work we consider the 3$d$ 3-state Potts model, which has an interesting phase diagram in the 
inverse temperature-magnetic field plane (see Fig.~\ref{phase_diag_xitau}) and provides a test-field
for universality ideas in two different regions: (a) near the critical endpoint in the $(\beta-h)$ plane, 
belonging to the 3$d$ Ising universality class~\cite{Karsch-Stickan} and (b) near the $h=0$ weakly 
first order transition in the broken phase.
In both cases we calculate the ratio between two massive excitations to be compared in the case (a) 
with the corresponding mass ratio of the 3$d$ Ising class; in the case (b) with the corresponding
ratio of screening masses in the (3+1)$d$ SU(3) pure gauge theory at finite temperature 
near the deconfinement transition in the broken phase, calculated in 
Ref.~\cite{Falcone-Fiore-Gravina-Papa_SU3}. The reason of the latter comparison is to check if and to 
what extent Svetitsky-Yaffe conjecture holds in the case of weakly first order transition. 

%%%%%%%%%%%%%%%%%%%%%%%%%%
%%%%%%%%%%%%%%%%%%%%%%%%%%

\section{The 3$d$ 3-state Potts model and massive excitations}
\label{Potts}

The 3$d$ 3-states Potts model~\cite{Blote-Swendsen,Janke-Villanova} is a spin 
theory in which the fundamental degree of freedom, $s_i$, defined in the site $i$ of a 
3-dimensional lattice, is an element of the Z(3) group, {\it i.e.}
\be
s_i=e^{i\frac{2}{3}\pi\sigma_i}, \quad  \sigma_i=\{0,1,2\}\quad .
\ee
The Hamiltonian of the model is
\be
H=-\frac{2}{3}\beta \sum_{\langle i j \rangle} \big( s_i^\dagger s_j 
+ s_j^\dagger s_i \big) =-\beta \sum_{\langle i j \rangle} \delta_{\sigma_i,\sigma_j}\;,
\label{ham}
\ee
up to an irrelevant constant. Here, $\beta$ is the coupling in units of the temperature and the sum is done
over all the nearest-neighbor pairs of a cubic lattice with linear size $L$. It is well known that this 
system undergoes a {\it weakly} first order phase transition~\cite{Gavai-Karsch-Petersson}, associated with 
the spontaneous breaking of the Z(3) symmetry. The order parameter of this transition is the magnetization, 
\be
\langle S \rangle=\langle \frac{1}{L^3}\sum_i s_i \rangle \;.
\ee
In presence of an external magnetic field it is convenient to work with an
Hamiltonian written in terms of the $\sigma_i$ degrees of freedom, instead of 
the $s_i$ ones. For a uniform magnetic field along the direction $\sigma_h$
with strength $h$ in units of the temperature, the Hamiltonian is
\be
H = -\beta \sum_{\langle i j \rangle} \delta_{\sigma_i,\sigma_j} 
-h \sum_i \delta_{\sigma_i,\sigma_h} \equiv -\beta E -h M \;,
\label{ham_magn}
\ee
where $E$ is the internal energy and $M$ is the magnetization.
The magnetic field breaks explicitly the Z(3) symmetry. However, first order 
transitions still occur for values of the magnetic field strength $h$ below a 
critical value $h_c$, the transition coupling decreasing with increasing $h$. 
The line of first order phase transitions ends in a second order critical point 
$P_c=(\beta_c,h_c)$ (see Fig.~\ref{phase_diag_xitau}), belonging to the 3$d$ Ising 
class~\cite{Karsch-Stickan}. The Hamiltonian in the scaling region 
of the critical point $P_c$ can be written as 
\be
H=-\tau \tilde{E} -\xi \tilde{M} \;,
\label{ising}
\ee
where $\tilde E$ and $\tilde M$ are the Ising-like energy and magnetization operators
and $\tau$ and $\xi$ the corresponding temperature-like and symmetry-breaking-like 
parameters. This means that $\langle \tilde M \rangle$ is the new order parameter.
Close enough to $P_c$, the following relations hold,
\be
\tilde{M}=M+sE \;, \;\;\;\;\; \tilde{E}=E+rM \;,
\label{mixing}
\ee
where the mixing parameters $(r,s)$ have been determined numerically for several 
lattice sizes $L$ in Ref.~\cite{Karsch-Stickan}. The $\tau$-direction identifies 
the first order line (see Fig.~\ref{phase_diag_xitau}).

Among the quantities relevant in the description of a phase transition there is
the correlation function of local order parameter: in our case of the 3$d$ 3-states 
Potts model this is just the local spin $s_i$. The point-point correlation
function is defined as
\be
\Gamma_{i_0}(r)=\langle s_i^\dagger s_{i_0} \rangle -\langle s_i^\dagger \rangle 
\langle s_{i_0} \rangle \;,
\ee
where $i$ and $i_0$ are the indices of sites and $r$ is the distance between them. 
The large-$r$ behavior of the point-point correlation function is determined by
the connected correlation length of the theory, $\xi_0$, or, equivalently, by
its inverse, the fundamental mass. It is convenient to study the connected wall-wall correlator in the 
$x$-direction defined as
\be
G(x)=\frac{1}{L}\sum_{x_0} \langle w(x_0+x)^\dagger w(x_0) \rangle -\langle 
w(x+x_0)^\dagger \rangle \langle w(x_0) \rangle \;,
\;\;\;\;\; w(x)=\frac{1}{L^2}\sum_{y,z} s_{\{x,y,z\}}
\label{corr}
\ee
where $w(x)$ represents the spin average over the ``wall'' at the coordinate $x$.

The general behavior for the function $G(x)$ is
\be
G(x)= \sum_n a_n e^{-m_n x} \;,
\label{corr-funct}
\ee
where $m_0$ is the fundamental mass, while $m_1$, $m_2$, ... are higher masses
with the same angular momentum and parity quantum numbers of the fundamental mass.
Mass excitations in channels different from 0$^+$ can be determined by a 
suitable redefinition of the wall average. The fundamental mass
in a definite channel can be extracted from wall-wall correlators by looking for 
a plateau of the effective mass, $m_{\mbox{\footnotesize eff}}(x)= -\ln G(x)/G(x-1)$,
at large distances.
In the present work we consider only the 0$^+$ and the 2$^+$ channels; the 
local variables to be wall-averaged as in~(\ref{corr}) have been defined 
in the following way:
\[
s^{0^+}_{\{x,y,z\}}(n)=s_{\{x,y,z\}}(s_{\{x,y+n,z\}}+s_{\{x,y,z+n\}}) \;,\;\;\;\;\;\; 
s^{2^+}_{\{x,y,z\}}(n)=s_{\{x,y,z\}}(s_{\{x,y+n,z\}}-s_{\{x,y,z+n\}}) \;.
\label{var}
\]

According to the conjecture of universal spectrum, ratios between massive excitations in 
the broken phase must be equal in theories belonging to the same universality class. The 
aim of this work is, then, to compare $m_{2^+}/m_{0^+}$ ratio in two cases: (a) near the 
3$d$ Ising critical endpoint to verify whether this is consistent with the class prediction;
(b) in the broken phase near the first order transition in absence of the magnetic field.
In the first case, the procedure to determine the fundamental masses in the
two channels of interest is the same outlined above, with the only 
difference that we need to use the correct local order parameter to build 
correlators. We have defined this local variable, $\tilde{m}_i$, in such a way that
it reconstructs the global magnetization operator 
$\tilde{M}$ after summation over the whole lattice:
\be
\tilde{m_i} = \delta_{\sigma_i \sigma_h} + \frac{s}{2} \sum_{\hat{\mu}} 
\delta_{\sigma_i \sigma_{i+\hat{\mu}}} \;.
\ee
Indeed, it is easy to see that $\tilde{M} = \sum_i \tilde{m}_i$.  
In the case (b), results are to be compared with the would-be universal partner of the Potts model, 
which is the (3+1)$d$ SU(3) pure gauge theory at finite temperature, to test if Svetitsky-Yaffe conjecture 
holds. Generally universality applies in case of critical transitions, {\it i.e.} where correlation length 
diverges. In the case of $h=0$ Potts model, as well as for SU(3) pure gauge theory, transition 
is weakly first order ($\xi$ keeping finite); this means that $\xi$ takes values much larger
than the lattice spacing, although finite.

\section{Numerical results}
\label{results}

\begin{figure}[tb]
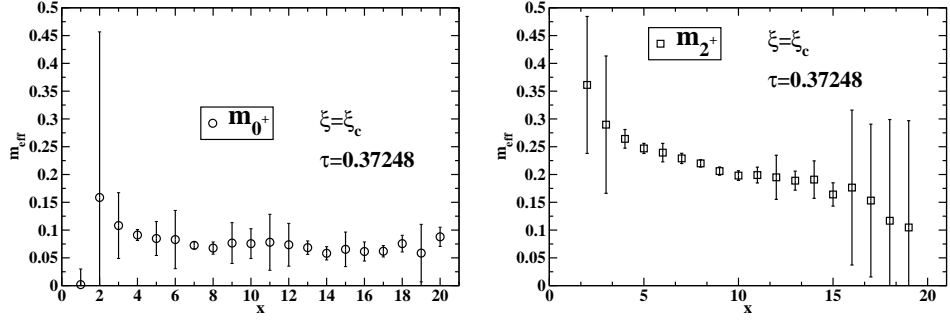

\centering
\includegraphics[width=6cm]{mass0+_xi=xi_c_tau=0.37248_left.eps} \quad \includegraphics[width=6cm]{mass2+_xi=xi_c_tau=0.37248_left.eps}
\caption[]{Effective mass in the 0$^+$ and 2$^+$ channels as a function of the separation between 
walls on the $(y,z)$ plane at $\xi=\xi_c$ and $\tau$=0.37248, determined from the 
configurations belonging to the ``right-peak'' in the thermal equilibrium ensemble.}
\label{karsch_mass0+_left} 
\end{figure}

We have performed numerical Monte Carlo simulations of the 3$d$ 3-states Potts 
model using a cluster algorithm~\cite{Swendsen-Wang,Kasteleyn-Fortuin} to reduce 
the autocorrelation effects. In order to minimize the finite volume effects, we have 
imposed periodic boundary conditions. Data analysis has been done by the jackknife
method applied to bins of different lengths. In both the regions studied we have seen tunneling between 
degenerate minima near the transition point. This finite volume effect can 
spoil mass measurements in the scaling region and must be treated carefully. 
Depending on the order of the transition, tunneling effects show up
differently and must be removed accordingly.
In the region (a) we have performed simulations on $70^3$ lattices for which the 
mixing parameters appearing in~(\ref{mixing}) turn to be $s(L=70)=-0.689(8)$ and 
$r(L=70)=0.690(3)$~\cite{Karsch-Stickan}. In this case tunneling occurs between two 
Z(2)-broken minima; this effect is removed simulating the system far enough ($\xi=\xi_c$,$\tau=0.37248$) 
from the transition, in the scaling region, where two corresponding Z(2)-peaks 
(``right'' and ``left'') are well separated. As an example, in Fig.~\ref{karsch_mass0+_left} 
it is displayed the behavior of the effective mass in the $0^+$ and $2^+$ channels for 
the data in the ``right'' Z(2)-peak; the fundamental masses in each channel turn to be 
the plateau values of the effective mass. We have found
\begin{eqnarray*}
\mbox{``right-peak'' (stat. 115K):} \;\;\;\;\;&
am_{0^+}= 0.0725(63) \;, \;\;\;
am_{2^+}= 0.1981(87) \;, \;\;\;
\frac{m_{2^+}}{m_{0^+}} = 2.73(36) \;; \\
\mbox{``left-peak'' (stat. 85K):} \;\;\;\;\; & 
am_{0^+}= 0.0714(40) \;, \;\;\;
am_{2^+}= 0.1959(80) \;, \;\;\;
\frac{m_{2^+}}{m_{0^+}} = 2.74(27) \;. 
\end{eqnarray*}
We can see that the mass ratios are consistent, as expected and, moreover, they are
compatible with the value of the 3$d$ Ising class~\cite{Caselle:1999tm}, 
$m_{2^+}/m_{0^+}= 2.59(4)$.

In the region (b) we have performed simulations on $48^3$ lattices for several values of the
coupling $\beta$ in the broken phase of the 3$d$ 3-state Potts model at zero magnetic
field. 

%%%%%%%%%%%%%%%%%%%%%%%%%%
%%%%%%%%%%%%%%%%%%%%%%%%%%

\begin{figure}[htb]
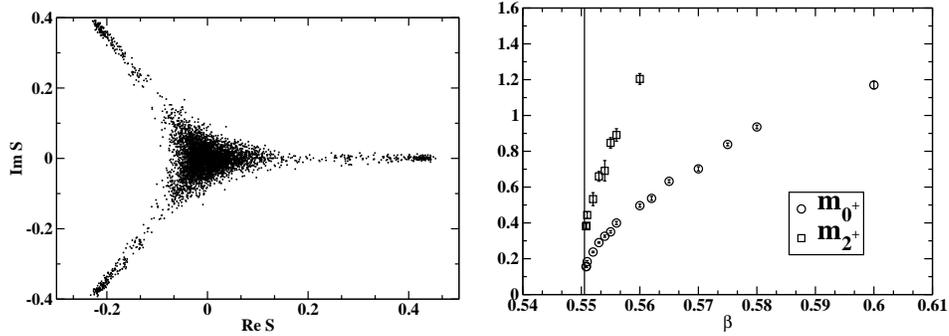

\centering
\includegraphics[width=6cm]{par0_norot_crit.eps} \quad \includegraphics[width=6cm]{mass0+_mass2+.eps}
\caption[]{(Left) Scatter plot of the complex order parameter $S$ at the transition
point $\beta_t$=0.550538~\cite{Gavai-Karsch-Petersson} at zero magnetic field. 
Both the symmetric and the broken phases are present. (Right) Fundamental masses 
in the $0^+$ and $2^+$ channels as functions of $\beta$, 
in the broken phase near $\beta_t$ (vertical line).}
\label{scatter_trans_masses_0_2}
\end{figure}

Close enough to $\beta_t(L=48)$=0.550538~\cite{Gavai-Karsch-Petersson}, the scatter plot of the complex order 
parameter $\langle S\rangle$ shows the coexistence of the symmetric phase 
(points around (0,0) in the Im$\langle S\rangle$ - Re$\langle S\rangle$ plane
in Fig.~\ref{scatter_trans_masses_0_2}) and of the broken phase (points around the three 
roots of the identity in Fig.~\ref{scatter_trans_masses_0_2}).

%%%%%%%%%%%%%%%%%%%%%%%%%%
%%%%%%%%%%%%%%%%%%%%%%%%%

\begin{figure}[htb]
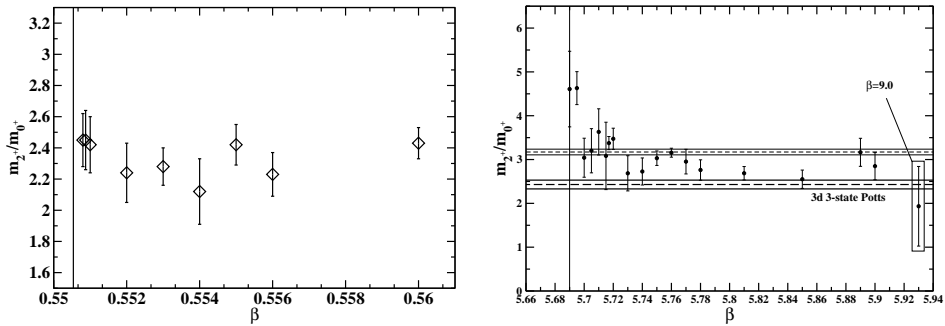

\centering
\bigskip
\includegraphics[width=6cm]{ratio_2_0.eps} \quad \includegraphics[width=6cm]{ratio_20_new.eps}
\caption[]{(Left) $m_{2^+}(\beta)/m_{0^+}(\beta)$ for $\beta$ varying in the scaling region. 
(Right) Ratio $m_{2^+}/m_{0^+}$ as a function of the coupling $\beta$ in the (3+1)$d$ SU(3) pure gauge 
theory near the deconfinement transition in the broken phase~\cite{Falcone-Fiore-Gravina-Papa_SU3}; 
the upper horizontal line is the fit of masses closest to transition and consistent with a constant; 
the lower one is the fit of the corresponding mass ratio in 3$d$ 3-state Potts model.}
\label{ratio_ratio_Po_SU3}
\end{figure}

To remove the former kind of tunneling we just moved away from the region across transition 
where symmetric phase appears; this happens up to $\beta=0.5508$. For the latter one, we 
applied an unambiguous rotation of all configurations belonging to the complex sectors to 
that along the real axis; this allows to improve statistics. We performed calculation of 
fundamental masses in $0^+$ and $2^+$ channels, as done in the case (a), up to $\beta=0.6$. 
In Fig.~\ref{scatter_trans_masses_0_2} we have plotted the behavior of $m_{0^+}$
and $m_{2^+}$ versus $\beta$ (for more details look at~\cite{Falcone-Fiore-Gravina-Papa}).
We have determined the ratio $m_{2^+}/m_{0^+}$ for several $\beta$ values
in the region $[\beta_t,0.56]$ which is plotted in Fig.~\ref{ratio_ratio_Po_SU3}. This 
ratio remains practically constant in the considered region, this suggesting that the 
correlation length $\xi_2$ associated to the channel $2^+$ ($\xi_2=1/m_{2^+}$) scales 
in the same way of the fundamental one ($\xi_0=1/m_{0^+}$). We can take as our estimation 
of the mass ratio the value
\be
\frac{m_{2^+}}{m_{0^+}} = 2.43(10) \;,
\label{ratioval}
\ee
determined by taking value and error of the point with the smallest error belonging to the plateau.
In Fig.~\ref{ratio_ratio_Po_SU3} we plotted the corresponding ratio in SU(3) pure gauge 
theory~\cite{Falcone-Fiore-Gravina-Papa_SU3} in the broken phase near the deconfinement 
transition and compared them with the result for the Potts model. As we can see the 
screening ratio turns to be larger than the corresponding one in the spin model.

%%%%%%%%%%%%%%%%%%%%%%%%%
%%%%%%%%%%%%%%%%%%%%%%%%%

%%%%%%%%%%%%%%%%%%%%%%%%%%
\section{Conclusions}

In this work we have studied massive excitations of the 3$d$ 3-states Potts model
near the Ising critical point on the inverse temperature - magnetic field phase diagram
and near the transition point at zero magnetic field.

We have found evidence that the mass ratio $m_{2^+}/m_{0^+}$ near the 
Ising critical point is compatible with the prediction from universality, thus
supporting the conjecture of universal spectrum.

In the broken phase of the scaling region near the transition in absence 
of the external source, we have found $m_{2^+}/m_{0^+}$=2.43(10).
This ratio turns to be $\simeq$30\% larger than the ratio of the lowest massive 
excitations in the same channels of the (3+1)$d$ SU(3) pure gauge theory at finite 
temperature in the broken phase. This can be taken as an estimate of the level of 
approximation by which the Svetitsky-Yaffe conjecture, valid in strict 
sense only for continuous phase transitions, can play some role also in this 
case of weakly first order transition.

\end{document}